\documentclass[onecolumn,showpacs,pra]{revtex4}
\usepackage{amsmath}
\usepackage{amsfonts}
\usepackage{amssymb}
\usepackage[cp1251]{inputenc}
\usepackage[english]{babel}
\usepackage{graphicx}% Include figure files
\usepackage{bm}% bold math

\newcommand{\ds}{\displaystyle}

\begin{document}

\title{FFLO states and quantum oscillations in mesoscopic superconductors
and superfluid ultracold Fermi gases}
\author{A.~V.~Samokhvalov$^{(1)}$,  A.~S.~Mel'nikov$^{(1)}$,
A.~I.~Buzdin$^{(2)}$}
\affiliation{$^{(1)}$ Institute for Physics
of Microstructures, Russian Academy
of Sciences, 603950 Nizhny Novgorod, GSP-105, Russia\\
$^{(2)}$ Institut Universitaire de France and Universit\'{e}
Bordeaux I, CPMOH, UMR 5798, 33405 Talence, France \\}


\date{\today}

\begin{abstract}
We have studied the distinctive features of the
Fulde-Ferrel-Larkin-Ovchinnikov (FFLO) instability and phase
transitions in two--dimensional (2D) mesoscopic superconductors
placed in magnetic field of arbitrary orientation and rotating
superfluid  Fermi gases with imbalanced state populations. Using a
generalized version of the phenomenological Ginzburg-Landau theory
we have shown that the FFLO states are  strongly modified by the
effect of the trapping potential confining the condensate. The
phenomenon of the inhomogeneous state formation is determined by
the interplay of three length scales: (i) length scale of the FFLO
instability; (ii) 2D system size; (iii)  length scale associated
with the orbital effect caused either by the Fermi condensate
rotation or magnetic field component applied perpendicular to the
superconducting disc. We have studied this interplay and resulting
quantum oscillation effects in both superconducting and superfluid
finite -- size systems with FFLO instability and described the
hallmarks of the FFLO phenomenon in a restricted geometry. The
finite size of the system is shown to affect strongly the
conditions of the observability of  switching between the states
with different vorticities.
\end{abstract}


\maketitle

\section{Introduction.}
The Zeeman interactions of electron spins with magnetic field is
known to be one of the mechanisms destroying the singlet
superconducting order (see, e.g., \cite{saint-james}). According
to this mechanism a homogeneous superconducting state becomes
energetically unfavorable above the Pauli limiting field
$H_p=\Delta/\mu_B\sqrt{2}$, where $\Delta$ is the gap value and
$\mu_B$ is the Bohr magneton. However, superconductivity can
appear even at the fields exceeding the $H_p$ field provided we
consider inhomogeneous states with a spatially modulated Cooper
pair wave function \cite{FFLO}. In this scenario the Cooper pairs
consist of electrons with different spin projections and different
absolute values of momentum.

There are at least two difficulties in experimental observation of
the FFLO instability: (i) first, the strong orbital effect which
destroys Cooper pairs above the upper critical field $H_{c2}$
which appears to be much less than $H_p$ in most superconducting
compounds; (ii) second, the impurity scattering which is known to
prevent the FFLO state formation. Thus, to observe this
interesting physical phenomenon we need to find rather clean
superconducting materials with very short coherence lengths to
increase the critical field corresponding to the orbital effect.
Alternatively, we should consider strongly anisotropic quasi two-dimensional (2D)
systems or very thin films in a magnetic field
parallel to the superconducting planes. Among the compounds which
are usually included in the list of strong candidates for the FFLO
states observation one should mention layered organic
superconductors \cite{organic}
and heavy fermion systems like $CeCoIn_5$ (see
\cite{h-fermion} and references therein).

During the last decade the attention of both theoreticians and
experimentalists have been attracted to a new type of superfluid
systems which are considered as promising playground for the study
of this intriguing phenomenon, i.e., ultracold Fermi gases in
magneto-optical traps \cite{fermion-gas}. The FFLO type
instability in these systems is caused not by the Zeeman
interaction but by the tuning of the population imbalance between
two lowest hyperfine states of $^6Li$ atoms. Experimentally this
population imbalance is governed by the radio frequency signal
inducing transitions between the hyperfine states. Thus, changing
the population imbalance we should get the inhomogeneous FFLO
state with a certain intrinsic length scale and this phenomenon is
not masked by any kind of the orbital effect. The orbital effect
in such neutral atomic condensates is associated not with magnetic
field but with system rotation which is known to be an important
part of the experimental procedure of detection of superfluidity
in the ultracold gases \cite{rotation}. The FFLO states in an
ultracold gas cloud should be, of course, modified by the effect
of the trapping potential confining the atomic system. As a
result, the physics of this phenomenon will be determined by the
interplay of three length scales: (i) length scale of the FFLO
instability; (ii) atomic system size; (iii) the length scale
associated with the condensate rotation
$L_\Omega=\sqrt{\hbar/M\Omega}$, where $M$ is the atomic mass and
$\Omega$ is the angular velocity. An analogous interplay appears
in a thin mesoscopic superconducting disc with FFLO instability
caused by the strong magnetic field parallel to the disc plane.
The effect of rotation in this case and corresponding length
$L_\Omega$ should be replaced by  the magnetic field component
$H_z$ perpendicular to the disc plane and magnetic length
$L_H=\sqrt{\hbar c/eH_z}$, respectively. The goal of this paper is
to study this interplay in both superconducting and superfluid
finite -- size systems  with FFLO instability and describe the
hallmarks of the FFLO phenomenon in a restricted geometry.

In  standard superconductors without FFLO instability the finite
system size is known to cause the so called Little -- Parks
effect, i.e. the oscillatory behavior of the phase transition line
on the plane magnetic field -- temperature \cite{LP,meso}. These
quantum oscillations originate from the switching between the
superconducting states with different vorticities or winding
numbers. The quantum oscillations of the critical temperature vs
magnetic field (or angular velocity) are known to reveal
themselves also in infinite 2D FFLO superconductors and
superfluids \cite{Buzdin-Kulic-JLTP84,Kulic}. Our theoretical work
aims to the identification of both the similarities and
distinctive features of the quantum oscillations in  mesoscopic
systems with and without FFLO instability. We focus here on the
case of 2D systems when the quantum oscillatory effects appear to
be most pronounced. In section II we discuss a modified Ginzburg
-- Landau model which takes account of both the FFLO phenomenon
and confinement effect. In section III we consider the case of a
mesoscopic disc while the section IV is devoted to the rotating
Fermi condensates confined in traps. We summarize our results in
section V.

\section{Modified Ginzburg -- Landau model for 2D FFLO states}

Hereafter our consideration of the FFLO phase formation will be
based on modified Ginzburg--Landay (GL) theory where the
appearance of the nonuniform state is caused by a change in the
sign of the second-order gradient term in the free energy
expansion. An appropriate GL functional can be derived from the
microscopic theory (see Ref.~\cite{buzdin-kach}). Calculating the
superfluid critical temperature one can take the GL free energy
density in the form:
\begin{equation} \label{eq:1v}
    F = a | \Psi |^2 - \beta | \mathbf{D} \Psi |^2
        + \gamma | \mathbf{D}^2 \Psi|^2\,.
\end{equation}
where $\Psi$ is the superfluid order parameter, $a=\alpha (T -
T_{c0})$ and $T_{c0}$ is the critical temperature of the
second--order transition into a uniform superconducting or
superfluid state, and $ \mathbf{D}$ is the gauge--invariant two
dimensional momentum operator. Note that here we omit the terms of
the higher order in  $\Psi$ which come into play only below the
superfluid transition. In the FFLO region the coefficients
$\beta,\,\gamma > 0\,$ and minimum of the free energy functional
does not correspond to uniform state, since a spatial variation of
the order parameter decreases of the energy of the system.
Certainly, the GL functional provides an adequate description of a
long-wavelength FFLO modulation only near the Lifshitz tricritical
point, however the results of the GL approach can be extrapolated
qualitatively to the whole region of the FFLO phase.

A lateral confinement of the condensate can be introduced either
using a boundary condition for the order parameter $\Psi$ at the
sample edge or adding an external potential well $V(\mathbf{r})$
to the free energy density (\ref{eq:1v}):
\begin{equation} \label{eq:2v}
    F = (a  + V(\mathbf{r})) |\Psi|^2 - \beta |\mathbf{D} \Psi|^2
        + \gamma |\mathbf{D}^2 \Psi|^2\,,
\end{equation}
where $\mathbf{r}$ is the in-plane radius vector. Varying an
appropriate free energy functional (\ref{eq:2v})  we find:
\begin{equation} \label{eq:3v}
    \gamma\,\mathbf{D}^4 \Psi + \beta\,\mathbf{D}^2 \Psi
       + \left( a + V(r) \right) \Psi = 0 \, .
\end{equation}
We restrict ourselves to the consideration of cylindrically
symmetric systems and, thus, assume the confining potential $V(r)$
to depend only on the radius $r$, where ($r$, $\theta$, $z$) are
the cylindrical coordinates. The value $k_0 = \sqrt{\beta
/2\gamma}$ in the above equation plays the role of the inverse
characteristic length scale of the FFLO modulation. It is
convenient to introduce a dimensionless coordinate $\rho = k_0 r$
and dimensionless shift of the critical temperature $\tau =
a/\gamma k_0^4$:
\begin{equation} \label{eq:6v}
    T_c = T_{c0} + \frac{\gamma k_0^4}{\alpha}\, \tau\, .
\end{equation}
As a result,  one can rewrite the the equation (\ref{eq:3v}) in a
dimensionless form:
\begin{equation} \label{eq:4v}
    \mathbf{D}^4_{\rho,\theta} \Psi + 2\,\mathbf{D}^2_{\rho,\theta} \Psi
       + \left( \tau + v(\rho) \right) \Psi = 0 \, ,
\end{equation}
where $\mathbf{D}_{\rho,\theta} = \mathbf{D} / k_0$ and  $v(\rho)
= V / \gamma\, k_0^4$.

In the following sections we proceed with the calculation of the
shift of the critical temperature for FFLO states with different
vorticities or winding numbers. We consider two generic examples
of restricted FFLO systems: (i) a thin mesoscopic superconducting
disk of the radius $R$ placed in an external magnetic field tilted
with respect to the disc plane; (ii)  rotating 2D superfluid Fermi
condensate confined in a harmonic trap.

%%%%%%%%%%%%%%%%%%%%%%%%%%%%%%%%%%%%%%%%%%%%%%%%%%%%%%%%%%%%%%%%%%%%
\begin{figure}[h]
\includegraphics[width=0.4\textwidth]{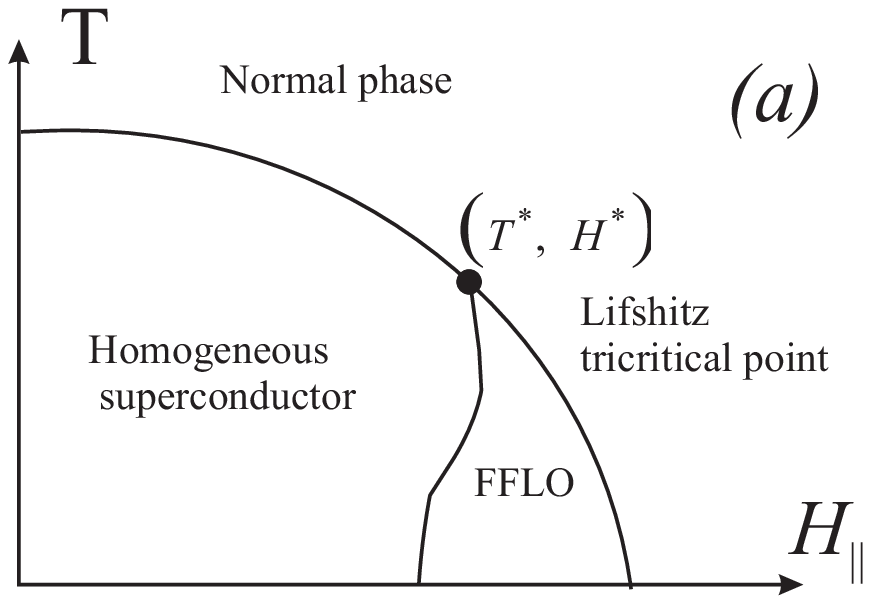}\hspace{30pt}
\includegraphics[width=0.4\textwidth]{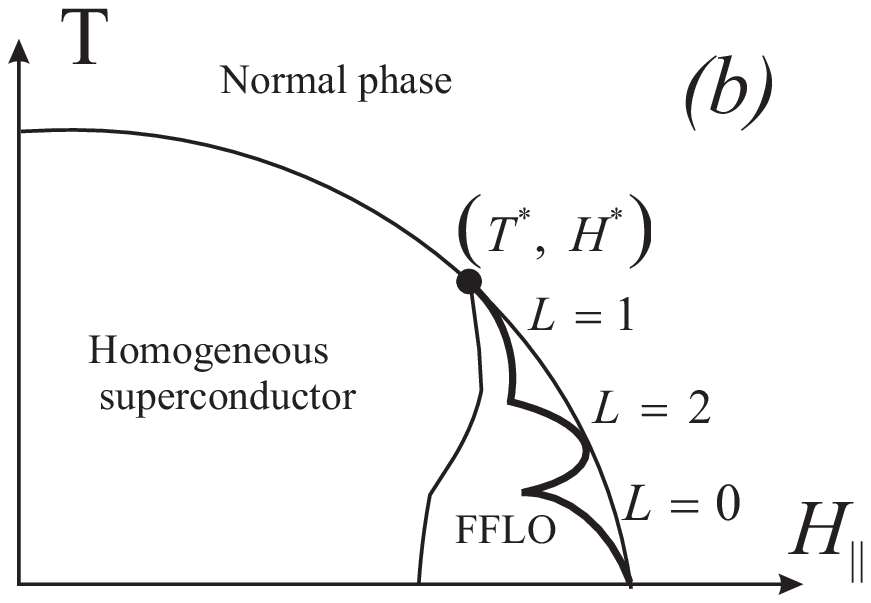}
\caption{Schematic phase diagram in the plane $H_\|-T$ for an
infinite 2D superconducting film (a) and for a 2D disc of a finite
radius (b) in a parallel magnetic field $H_\|$.} \label{Fig:1}
\end{figure}
%%%%%%%%%%%%%%%%%%%%%%%%%%%%%%%%%%%%%%%%%%%%%%%%%%%%%%%%%%%%%%%%%%%%
%

\section{FFLO state in a 2D mesoscopic disc}

A thin superconducting disc of a finite radius $R$ placed in
external magnetic field $\mathbf{H} = \mathbf{H}_{\parallel} + H_z
\mathbf{z}_0$ provides a simplest example illustrating the effect
of Cooper pair confinement on the FFLO state. The gauge--invariant
2D momentum operator in the above equations for $\Psi$ takes the
form:
$$
    \mathbf{D} = \nabla + \frac{2\pi i}{\phi_0}\, \mathbf{A}\ ,
$$
where $\mathbf{A}= (0\,, A_\theta\,, 0)= (0\,, H_z r / 2, 0)$ is
the vector--potential of the field component $H_z =
\mathrm{curl}_z \mathbf{A}$, and $\phi_0 = \pi \hbar c / | e |$ is
the flux quantum. Considering the limit of vanishing disc
thickness we neglect here the orbital effect caused by the field
component $\mathbf{H}_{\parallel}$. At the same time the Zeeman
interaction energy associated with this parallel field component
$\mathbf{H}_{\parallel}$ is assumed to be crucial and responsible
for the FFLO instability. The coefficient $\beta=\beta(H_\|\,,T)$
is a function of temperature $T$ and Zeeman energy $\mu_B H_\|$
and vanishes in the tricritical Lifshitz point
($T^*,\,H^*=H_{c2}(T^*)$) : $\beta(H^*, T^*)=0$. This tricritical
point $(T^*,H^*)$ is the meeting point of three transitions lines
separating the normal, uniform superconducting and nonuniform FFLO
states (see Fig.~1a).
A trapping potential is assumed to be absent
($V(r) = 0$) and confinement of the superconducting condensate
occurs due to the boundary condition at the disc edge. This
Neumann-type boundary condition for a disc in an insulating
environment and the gauge  $\mathbf{A}= (0\,, A_\theta\,, 0)$
takes the form:
\begin{equation} \nonumber
    \left.\frac{\partial \Psi}{\partial r} \right|_{r = R} = 0\,.
\end{equation}

\subsection{FFLO state in a 2D mesoscopic disc in a parallel
magnetic field.}

We start our consideration from the case of zero perpendicular
component of external magnetic field: $H_z=0$. The equation
(\ref{eq:4v})
 can be simplified and written as follows:
\begin{equation} \label{eq:8v}
    \Delta^2_{\rho,\theta} \Psi + 2\,\Delta_{\rho,\theta} \Psi + \tau \Psi =
    0\, ,
\end{equation}
where $\Delta_{\rho,\theta}$ is a 2D Laplace operator written in
dimensionless coordinates $\rho, \theta$. The equation
(\ref{eq:8v}) with the boundary condition
\begin{equation} \label{eq:7v}
    \left.\frac{\partial \Psi}{\partial \rho} \right|_{\rho = R_0} = 0\,
\end{equation}
defines a set of eigenfunctions and corresponding eigenvalues $\tau$.
Here we introduce the dimensionless disc radius $R_0 = k_0 R$.
The maximum eigenvalue $\tau$ gives us a critical temperature of
transition into the FFLO phase. The solution can be simplified due
to the following obvious observation: the eigenfunctions of the
equation (\ref{eq:8v}) coincide with eigenfunctions of the
Schr\"odinger -- like problem
\begin{equation} \label{eq:9v}
    -\Delta_{\rho,\theta} \Psi = q^2 \Psi \ ,
\end{equation}
with the boundary condition (\ref{eq:7v}) at the disc edge. The
resulting dimensionless shift of the critical temperature $\tau$
depends on the wave number $q$:
\begin{equation} \label{eq:10v}
    \tau(q) = 2 q^2 - q^4 \ .
\end{equation}
The solutions of Eq.~(\ref{eq:9v}) characterized by a certain
angular momentum $L$ can be expressed via the Bessel function of
first kind $J_L (q \rho)$:
\begin{equation} \label{eq:11v}
    \Psi = \mathrm{e}^{i L \theta} J_L (q \rho)\,.
\end{equation}
The vorticity  parameter $L$ coincides with the angular momentum
of the Cooper pair wave function. The boundary condition
(\ref{eq:7v}) gives us a set of zeros $z_{L n}$ of the derivative
of the Bessel function $J_L (z)$: $\partial_z J_L (z_{L n}) = 0$.
As a consequence, we get a set of eigenvalues $q_{L n} = z_{L n} /
R_0$. In accordance with (\ref{eq:10v}) the set of wave numbers
$q_{L n}$ determines  a set of critical temperature shifts
\begin{equation} \label{eq:12v}
   \tau_{L n} = 2 \left( \frac{z_{L n}}{R_0} \right)^2
             - \left( \frac{z_{L n}}{R_0} \right)^4 \,,
\end{equation}
characterizing vortex states with different winding numbers $L$:
$$
    \Psi_{L n} = \mathrm{e}^{i L \theta} J_L (q_{L n} \rho) \ .
$$
To get the critical temperature of the superconducting transition
into the FFLO state we need to find the maximum of the $T_c$
value, i.e. the maximum of the function
\begin{equation} \label{eq:18}
     T_c -T_{c0}  = \frac{\gamma k_0^4}{\alpha}\, \underset{L\, n}{\rm max}\{\tau_{L n}\}\,.
\end{equation}

In Fig.~2 we plot the dependencies of the dimensionless shift of
the critical temperature $\tau_{L n}$ vs the parameter $R_0$ for
different $L$ and $n$ values. For the fixed value of the disk
radius $R$ the parameter $R_0$ can be tuned by changing  the
temperature $T$ and/or the in-plane magnetic field $H_\|$. We see
that for a small disk radius $R \ll 1 / k_0$ ($R_0 \ll 1$) FFLO
instability is suppressed ($\tau < 0$) and only uniform
superconducting state appears to be energetically favorable. With
the increase in the $R_0$ value the diameter of the disc becomes
comparable with the period of the superconducting order parameter
oscillations and, thus, FFLO state in the disk becomes
energetically favorable. It is interesting to note that nonuniform
FFLO state promotes the vortex states with $L \ne 0$: the mode
$L=1$ arises primarily just below $T^*$. The switching between the
FFLO states characterized by different winding numbers $L$ results
in an oscillatory behavior of the critical temperature $T_c$ as a
function of the external field $H_\|$. In Fig.~1b we show
schematically a typical phase diagram in the plane $H_\|-T$. The
critical temperature appears to be degenerate for FFLO states with
opposite vorticity signs and, as a result, the
sinusoidally modulated superconducting
states below $T_c$ can be formed by superpositions of angular
harmonics with $L$ and $-L$ similar to those observed numerically
in mesoscopic rings \cite{fei}.
%
%%%%%%%%%%%%%%%%%%%%%%%%%%%%%%%%%%%%%%%%%%%%%%%%%%%%%%%%%%%%%%%%%%%%
\begin{figure}%[t]
\includegraphics[width=0.55\textwidth]{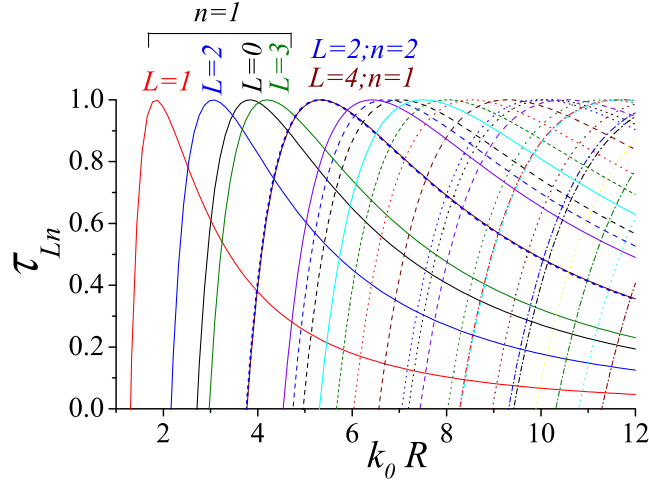}
\caption{(Color online) Dependence of the shift of the critical temperature
$\tau$ vs the dimensionless disc radius $k_0R$ for different
values of vorticity $L$.} \label{Fig:2}
\end{figure}
%%%%%%%%%%%%%%%%%%%%%%%%%%%%%%%%%%%%%%%%%%%%%%%%%%%%%%%%%%%%%%%%%%%%

\subsection{FFLO state in a mesoscopic disc in the magnetic field
of arbitrary orientation.  Little-Parks Oscillations.}

Let us now consider the effect of an additional component of the
magnetic field $H_z$, applied perpendicular to the disc plane. We
use here the gauge $\mathbf{A} = (0\,, A_\theta\,, 0)$ where
$A_\theta = H_z r / 2$, and look for the solution of the
Eq.~(\ref{eq:4v}) (with $v = 0$) characterized by certain angular
momentum $L$
\begin{equation} \label{eq:14v}
    \Psi(\rho,\theta) = f_L(\rho)\,\mathrm{ e}^{iL\theta}\, .
\end{equation}
The function $f_L(\rho)$ satisfies the equation
\begin{equation} \label{eq:15v}
    \mathbf{D}_L^2\left(\mathbf{D}_L^2\, f_L \right) + 2 \mathbf{D}_L^2\, f_L
    + \tau \, f_L = 0 \, ,
\end{equation}
where the operator $\mathbf{D}_L$ is determined by the expression
\begin{equation} \label{eq:16v}
    \mathbf{D}_L^2 = \frac{1}{\rho}\frac{d}{d \rho}
            \left( \rho \frac{d}{d \rho} \right)
           - \left( \frac{L}{\rho} + \frac{\rho}{a^2_H} \right)^2\,.
\end{equation}
Here $a_H = k_0 \sqrt{\phi_0 / \pi H_z}$ is the dimensionless
magnetic length in the units of $k^{-1}_0$. The solution in the
disc should meet the boundary condition
\begin{equation} \label{eq:17v}
    \frac{\partial f_L}{\partial \rho}\,{\bigg\vert_{\ds \rho = R_0}} = 0
\end{equation}
at the disk edge. As in the previous subsection, the eigenvalue
$\tau$ determines the shift of the critical temperature caused by
the  FFLO instability. The eigenfunctions $f_L(\rho)$ of the
problem (\ref{eq:15v}), (\ref{eq:17v}) coincide with the
eigenfunctions of the differential operator $\mathbf{D}_L^2$
\begin{equation} \label{eq:18v}
    - \mathbf{D}_L^2 f_L =  q^2 f_L \,,
\end{equation}
with the same boundary condition (\ref{eq:17v}). The relation
between the eigenvalue $\tau$ and the eigenvalue of the operator
$\mathbf{D}_L^2$ is given by the expression (\ref{eq:10v}).

The solution of the equation (\ref{eq:18v}) can be expressed via
the confluent hypergeometric function of the first kind (Kummer's
function) $F(a,b,z)$ \cite{Abramowitz-Stegun}
\begin{equation} \label{eq:19v}
    f_L(\phi)= \mathrm{e}^{-\phi/2} \phi^{\vert\,L\,\vert/2}
             F\left(\,a_L,\,b_L,\,\phi\,\right)\,,
\end{equation}
where
\begin{equation} \label{eq:20v}
    a_L = \frac{1}{2}\left( \vert L \vert + L + 1
        -\frac{q^2 a^2_H}{2} \right)\,,
\quad
    b_L = \vert L \vert +1\, ,
\quad
    \phi = \rho^2 / a^2_H\, .
\end{equation}
The boundary condition (\ref{eq:17v}) can be rewritten in terms of the Kummer's
functions:
\begin{equation} \label{eq:21v}
    a_L F\left( a_L+1, b_L+1, \phi_R \right)
    + \frac{b_L}{2}\left(\frac{| L |}{\phi_R} - 1 \right)
            F( a_L, b_L, \phi_R) = 0 \ ,
\end{equation}
where $\phi_R = \pi R^2 H_z / \phi_0$  is the magnetic flux
piercing the disk area in the units of flux quantum. The equations
(\ref{eq:20v}) and (\ref{eq:21v}) define an implicit dependence of
the eigenvalue $q_L$ on the parameters $k_0$, $R$, $H_z$ and the
orbital number $L$. Thus, using Eq.~(\ref{eq:10v}) one obtains the
dependence of the critical temperature $T_L$ of the state with a
vorticity $L$ on the parameters $k_0$, $R$, $H_z$:
\begin{equation} \label{eq:23v}
    T_L = T_{c0} + \frac{\gamma k_0^4}{\alpha}\, \tau_L \, ,
\end{equation}
where
\begin{equation} \label{eq:22v}
    \tau_L = 2 q_L^2 - q_L^4\,.
\end{equation}
The critical temperature $T_c$ of superconductivity nucleation  is
determined by the maximal $T_L$ value:
\begin{equation} \label{eq:18x}
    T_c = \underset{L}{\rm max}\{T_L\}\,.
\end{equation}
 The maximal $T_{L}$ corresponds to the maximal eigenvalue $\tau$
of the  problem (\ref{eq:15v}), (\ref{eq:17v}). It has been
already shown that for $H_z = 0$ the function $\tau(q)$ can be
expressed through the zeros of the derivatives of the Bessel
functions. These values were taken as the zero approximations to
roots of the general boundary condition (\ref{eq:21v}) for $H_z
\neq 0$.

In Figs.~3 and 4 we show typical phase diagrams on the plane
($\tau\,, \phi_a=1/a_H^2$) for different disk radii. Here $\phi_a$
is a dimensionless magnetic field component along the $z$ axis.
The phase boundary exhibits Little-Parks oscillations, caused by
transitions between the states with different angular momenta $L$.
For rather small disk radii (Fig.~3) one can clearly observe the
regime of the magnetic field induced superconductivity. It should
be noted that the switching between the vortex states in the disk
can occur with large jumps in vorticity $\triangle L > 1$ (see
Fig.~4b). Similar jumps in vorticity are known to occur in
mesoscopic rings \cite{Zyuzin} and hybrid FS structures
\cite{SMB}.

%%%%%%%%%%%%%%%%%%%%%%%%%%%%%%%%%%%%%%%%%%%%%
\begin{figure}[t]
\includegraphics[width=0.6\textwidth]{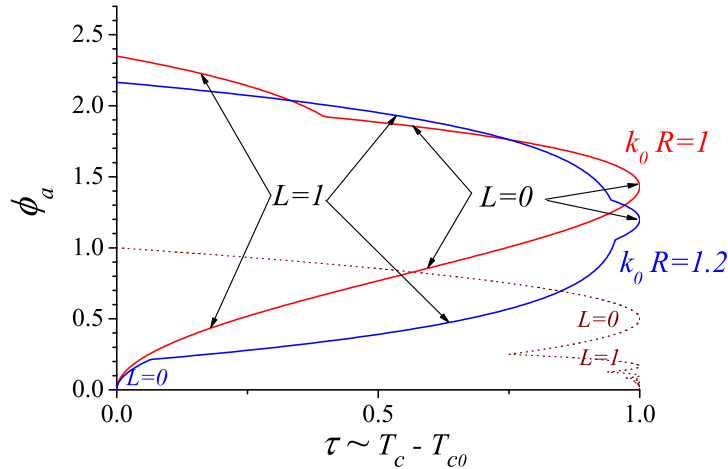}
\caption{(Color online) Typical phase diagrams for 2D discs in the
plane $(\tau, \phi_a)$ for disk radii  $k_0 R = 1\,,\: 1.2$. The
arrows point to the segments of the $\tau(\phi_a)$ curves
corresponding to different values of  vorticity $L=0, 1$. The
dotted line corresponding to an infinite 2D system (see
\cite{Buzdin-Kulic-JLTP84}) is shown for comparison.}
\end{figure}
%%%%%%%%%%%%%%%%%%%%%%%%%%%%%%%%%%%%%
%
%%%%%%%%%%%%%%%%%%%%%%%%%%%%%%%%%%%%%%%%%%%%%
\begin{figure}%[b]
\includegraphics[width=0.45\textwidth]{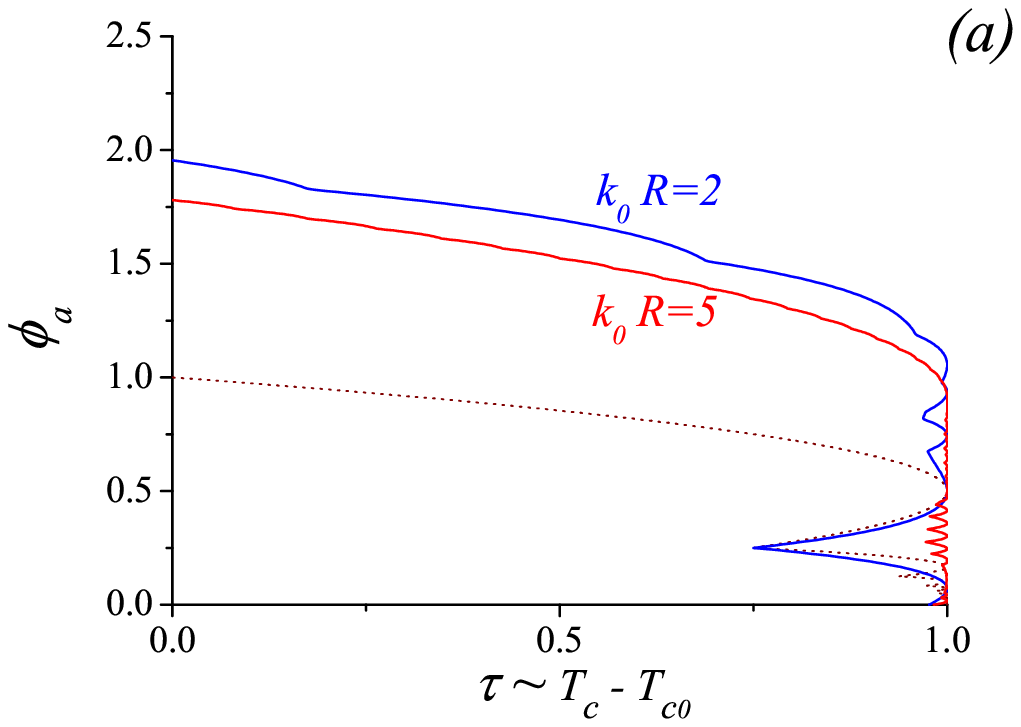}
\includegraphics[width=0.450\textwidth]{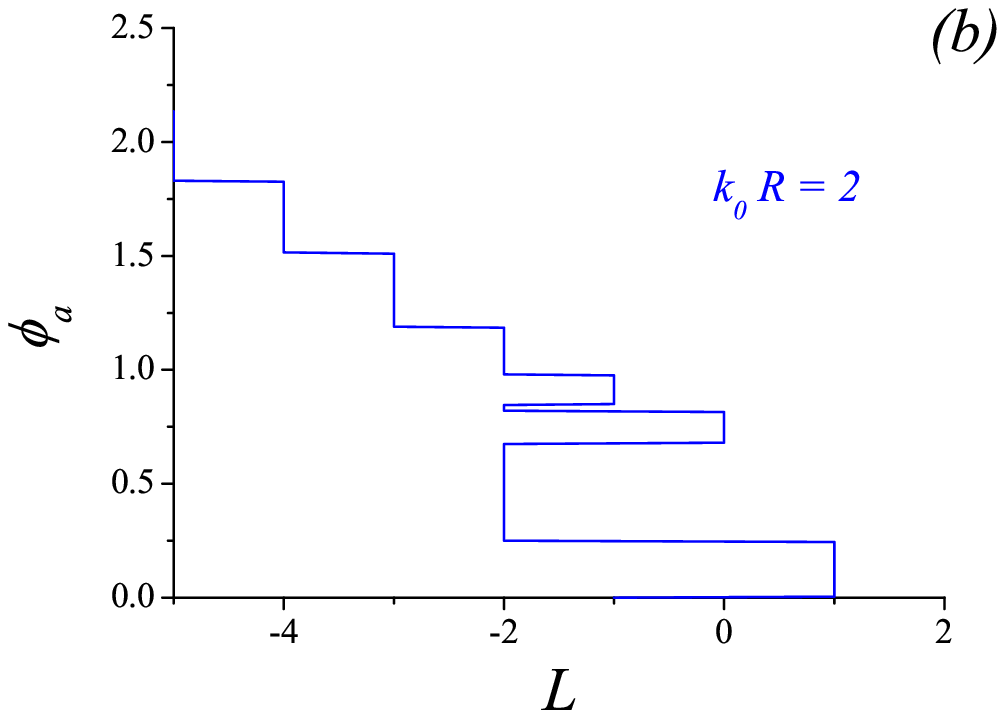}
\caption{(Color online) (a) Typical phase diagrams for 2D discs in the plane
$(\tau, \phi_a)$ for disk radii  $k_0 R = 2\,,\: 5$. The dotted
line corresponding to an infinite 2D system (see
\cite{Buzdin-Kulic-JLTP84}) is shown for comparison. (b) Jumps in
vorticity $L$ vs the dimensionless magnetic field $\phi_a$ for
$k_0 R=2$.}
\end{figure}
%%%%%%%%%%%%%%%%%%%%%%%%%%%%%%%%%%%%%

%\vspace{2cm}

\subsection{Vortex solution in a disc beyond the range of FFLO
instability.  Critical field of the vortex entry.}

Unconventional behavior of the vortex states in thin discs placed
in a strong parallel magnetic fields reveals, of course, not only
in the peculiarities of the oscillatory behavior of the
superconducting phase transition line. To illustrate the effect of
Zeeman interaction energy on the basic vortex matter properties in
finite size samples we consider here the critical field of the
first vortex entry into a homogeneous superconducting state close
(but beyond) the range of FFLO instability. In order to find this
critical field we need to calculate the energy difference between
the states with and without vortex. Neglecting the contribution of
the vortex core we can assume the order parameter absolute value
to be homogeneous ($\Psi\simeq e^{-i\theta}$) and consider only the
gradient part of the free energy functional:
\begin{equation}
    \frac{F_g}{F_0} = \int d^2 r\left(\xi_1^2 | \mathbf{D} \Psi |^2
        + \xi_2^4 | \mathbf{D}^2 \Psi|^2
        \right) \ ,
\end{equation}
where $F_0$ is a constant normalization factor. Approaching the
tricritical point one can change the balance between two gradient
terms in the above expression: for $H_\|\rightarrow H^*$ we obtain
$\xi_2\gg\xi_1$ and, thus, the fourth -- order gradient term
becomes a dominant one.

 We consider a vortex placed in the center of a disk of finite radius $R$
and take the gauge $A_\theta=H_z r/2$. The energy difference
between the states with and without such vortex takes the form:
$$
    \frac{\delta F_g}{F_0} = 2\pi\int\limits_{\xi_m}^R r d r\left[ \xi_1^2
    \left(\frac{1}{r} - \frac{2\pi}{\phi_0} A_\theta \right)^2
    - \xi_1^2 \left(\frac{2\pi}{\phi_0} A_\theta \right)^2
        + \xi_2^4 \left(\frac{1}{r} - \frac{2\pi}{\phi_0} A_\theta \right)^4
        - \xi_2^4 \left(\frac{2\pi}{\phi_0} A_\theta \right)^4
        \right] \ ,
$$
where $\xi_m = max[\xi_1, \xi_2]$. Integrating over $r$ we find:
$$
    \frac{\delta F_g}{F_0} = 2\pi \xi_1^2 \left[ \ln \frac{R}{\xi_m} - \phi_R
          + \frac{\xi_2^4}{2 \xi_m^2 \xi_1^2}
          + \frac{\xi_2^4}{R^2 \xi_1^2}
                \left(3\phi_R^2 -  \phi_R^3 - 4\phi_R \ln \frac{R}{\xi_m} \right)
        \right] \,,
$$
where $\phi_R=\pi R^2 H_z /\phi_0$. The condition $\delta F =0$ gives
us the field of first vortex entry:
%$$
%\ln \frac{R}{\xi_m} -\Phi +\frac{\xi_2^4}{2\xi_m^2\xi_1^2}
%        + \frac{\xi_2^4}{R^2\xi_1^2} \left(3\Phi^2
%-4\Phi \ln \frac{R}{\xi_m} -  \Phi^3
%         \right) =0
%         \ .
%$$
%We introduce $\tilde R = R/\xi_m\gg 1$ and $\alpha = \xi_2^4\xi_m^{-2}\xi_1^{-2}$:
$$
\ln \tilde R -\Phi +\frac{\alpha }{2} + \frac{\alpha}{\tilde R^2} \left(3\Phi^2
-4\Phi \ln \tilde R -  \Phi^3 \right) = 0  \,.
$$
Here we introduce the dimensionless parameters: $\tilde R = R/\xi_m\gg 1$ and
$\alpha = \xi_2^4 / (\xi_m \xi_1)^2$.
It is natural to consider now two limiting cases. Far from the
range of FFLO instability we can put $\xi_1\gg\xi_2$ ($\alpha\ll
1$) and find a standard logarithmic expression: $\phi_R\simeq \ln
(R/\xi_1)$. The field of the first vortex entry can be written as
follows:
$$
H^{(c)}_z\propto\frac{1}{R^2}\ln\frac{R}{\xi_1}\,, \qquad  \xi_1 \gg \xi_2\,.
$$
Close to the range of FFLO instability we need to consider an
opposite limit $\xi_1\ll\xi_2$ ($\alpha\gg 1$) and  obtain:
$\phi_R\simeq   (R/\xi_2) ^{2/3}$. The scaling behavior of the
field of the first vortex entry changes dramatically:
$$
    H^{(c)}_z\propto\left(\frac{1}{R}\right)^{4/3}\,,
    \qquad  \xi_2 \gg \xi_1\ .
$$
Both coherence lengths ($\xi_1$ and $\xi_2$) diverge as one
approaches the tricritical point. Considering the above
asymptotical expressions for $H^{(c)}_z$ one can see that for
$H_\|$ well below $H^*$ the field $H^{(c)}_z$ diverges as a
function of variable $H^*-H_\|$, while close to $H^*$ the critical
field $H^{(c)}_z$ tends to zero. Thus, the dependence of the
critical field $H^{(c)}_z$ vs $H^*-H_\|$ should reveal a peak in
the vicinity of the Lifshitz tricritical point.

%%%%%%%%%%%%%%%%%%%%%%%%%%%%%%%%%%%%%%%%%%%%%%%%%%%%%%%%%%%%%%%%%%%

\section{FFLO state in a superfluid condensate confined in a trap.}

As a second example of the effect of the condensate confinement on
the FFLO states  we consider a superfluid Fermi gas trapped by a
harmonic potential
\begin{equation} \label{eq:1g}
    V(r) = \frac{1}{2}\,M\omega^2 r^2\,.
\end{equation}
Here $\omega$ is a trapping frequency, $M$ is the atomic mass.
Similarly to the previous section we start from the free energy
density (\ref{eq:2v}) written in notations which are adequate for
a rotating superfluid gas. In this case the two dimensional
momentum operator $\mathbf{D}$ can be expressed through the
angular velocity vector $\mathbf{\Omega} = \Omega\, \mathbf{z}_0$
directed along the $z$ axis
$$
    \mathbf{D} = \nabla - \frac{2iM }{\hbar}\, [{\bf\Omega,{\bf r}}]\,,
$$
and the coefficient $\beta$ of the term $\beta |\nabla \Psi|^2$ in
the expression (\ref{eq:2v}) depends on the population imbalance
$\delta\mu$. The rotation of superfluid gases plays a similar role
as the orbital effect in superconductors. Varying the free energy
functional  and introducing a dimensionless radial coordinate
$\rho = k_0 r$ we find:
\begin{equation} \label{eq:2g}
    \mathbf{D}^4_{\rho,\theta} \Psi + 2\,\mathbf{D}^2_{\rho,\theta} \Psi
       + \left( \tau + v_0 \rho^2 \right) \Psi = 0 \, ,
\end{equation}
where $\mathbf{D}_{\rho,\theta}=\mathbf{D}/k_0$ and the parameter
$v_0 = M\omega^2 /2 \gamma k_0^6$ characterizes the trapping
potential.

\subsection{FFLO state in a parabolic trapping potential.}

In the absence of rotation ($\Omega = 0$) the Eq.~(\ref{eq:2g})
can be simplified:
\begin{equation} \label{eq:4g}
   \Delta^{2}_{\rho,\theta} \Psi + 2\Delta_{\rho,\theta} \Psi
       + ( \tau + v_0 \rho^2)\, \Psi = 0\, ,
\end{equation}
where $\Delta_{\rho,\theta}$ is a 2D Laplace operator written in
$\rho, \theta$ coordinates. Introducing a 2D Fourier transform
\begin{equation} \label{eq:5g}
\Psi = \int d^2 \mathbf{q}\, e^{i\,\mathbf{q}\, \mathbf{r}^\prime}\, \psi(\mathbf{q})
\end{equation}
one can write the equation (\ref{eq:4g}) in the momentum
representation as the Schr\"odinger--like equation with the
potential  $U(q)=q^4-2q^2$:
\begin{equation} \label{eq:6g}
-v_0\,\Delta_\mathbf{q}\,\psi + U(q)\,\psi = - \tau\psi \, .
\end{equation}
One can see that the solution of Eq.~(\ref{eq:4g}) with minimal
energy $-\tau$ should correspond to the zero angular momentum:
$L=0$. Indeed, the momentum dependent contribution to energy is
positive and proportional to $L^2$. For rather small $v_0$ values
the lowest energy level $-\tau$ in this Schr\"odinger -- like
equation is close to the value $-1$ and the wave function is
localized near the potential minimum. As a result, one can can
introduce the coordinate $s=q-1$ and expand the potential near the
minimum $U\simeq -1+4s^2$ to consider an approximate oscillator --
type solution. Indeed, for $|s|\ll 1$ we obtain:
\begin{equation} \label{eq:7g}
-v_0\frac{\partial^2}{\partial s^2}\psi + 4s^2\psi = (1-\tau)\psi\, .
\end{equation}
The lowest energy level of this harmonic oscillator and
corresponding wave function take the form:
$$
    \tau =1-2\sqrt{v_0} \ , \qquad
    \psi = e^{-s^2/\sqrt{v_0}} \ .
$$
The expression $\tau =1 - 2\sqrt{v_0}$ gives us the critical
temperature of the FFLO state. One can see that the FFLO
instabitily appears only for rather small trapping frequencies:
$v_0<1/4$. To find the eigenfunction in the ${\bf r}$ space we
should consider the inverse Fourier transform:
$$
\Psi \simeq \int\limits_{-\infty}^{+\infty} ds e^{-s^2/\sqrt{v_0}}
J_0((1+s)\rho) \ ,
$$
where $J_0$ is a Bessel function of the zeroth order. Considering
the asymptotical expression for the Bessel function at $\rho\gg 1$
we find:
$$
\Psi \simeq \frac{1}{\sqrt{\rho}}\int\limits_{-\infty}^{+\infty}
ds e^{-s^2/\sqrt{v_0}} \cos((1+s)\rho-\pi/4)
=\frac{1}{\sqrt{\rho}} Re \int\limits_{-\infty}^{+\infty} ds
e^{-s^2/\sqrt{v_0}} e^{i(1+s)\rho-i\pi/4}=
\sqrt{\frac{\pi}{\rho}}\cos(\rho-\pi/4)e^{-\rho^2\sqrt{v_0}/4}\ .
$$
Thus, the wave function strongly decays with increase in the
trapping frequency and the number of observable oscillations is of
the order of $2v_0^{-1/4} = 2 k_0 (\beta / M \omega^2)^{-1/4}$.

\subsection{FFLO states in a rotating superfluid gas in a parabolic
trapping potential.}

As a next step we study the effect of rotation ($\Omega \ne 0$) on
the superfluid states of the Fermi gas trapped in the parabolic
potential well (\ref{eq:1g}).  We look for the solution of
Eq.~(\ref{eq:2g}) characterized by the conserving angular momentum
$L$:
\begin{equation} \label{eq:8g}
    \Psi_L(\rho, \theta) = f_L(\rho)\,\mathrm{ e}^{i L \theta}\,,
\end{equation}
where $f_L$ satisfies the equation
\begin{equation} \label{eq:9g}
    \mathbf{D}_L^2\left(\mathbf{D}_L^2\, f_L \right) + 2 \mathbf{D}_L^2\, f_L
    + \left(\tau + v_0 \rho^2 \right)\, f_L = 0 \, ,
\end{equation}
\begin{equation} \label{eq:10g}
    \mathbf{D}_L^2 = \frac{1}{\rho}\frac{d}{d \rho}
            \left( \rho \frac{d}{d \rho} \right)
           - \left( \frac{L}{\rho} + \phi_a\rho \right)^2,\,
\end{equation}
and $\phi_a = 2M\Omega/\hbar k_0^2$ is the dimensionless rotation
frequency. Let us consider the following expansion for the order
parameter:
\begin{equation} \label{eq:11g}
    f_L(\rho) = \sum\limits_{n=0}^\infty c_n\, u_{n L}(\rho)\,,
\end{equation}
where $u_{nL}$ are the eigenfunctions of the operator $-{\bf
D}_L^2$ corresponding to the eigenvalues
$$
q_{nL}^2=2\phi_a(2n+L+|L|+1) \ ,
$$
and the coefficients $c_n$ satisfy the equation
\begin{equation} \label{eq:12g}
    \left( 2\,q_{nL}^2 - q_{nL}^4 \right)\, c_n
         - \sum\limits_m v_{n m}\, c_m = \tau\, c_n \,.
\end{equation}
The matrix elements
$$
    v_{nm}^L = v_0\, \int\limits_0^\infty\,
        \rho\,d\rho\,\left( u_{mL}\,\rho^2\, u_{n L} \right)
$$
are nonzero if $m=n$ or $m=n \pm 1$:
\begin{equation} \label{eq:13g}
    v_{n n}^L = \frac{v_0}{\phi_a} \left(2 n + \vert L \vert + 1 \right)\,, \quad
    v_{n (n+1)}^L = -\frac{v_0}{\phi_a} \sqrt{(n + 1)\,(n + \vert L \vert + 1)}\,, \quad
    v_{(n-1) n}^L = -\frac{v_0}{\phi_a} \sqrt{n\,(n + \vert L \vert)}\,.
\end{equation}
The set of normalized eigenfunctions $u_{n L}(\rho)$ can be
written as follows:
\begin{equation} \label{eq:14g}
    u_{n L}(\rho) = \sqrt{ 2 \phi_a\frac{(n + \vert L \vert)!}
                                 {n!(\vert L \vert)!)^2} }\,
                    \mathrm{e}^{-\phi_a \rho^2/2}
                    \left(\phi_a \rho^2\right)^{\vert\,L\,\vert / 2}
                        F\left(\,-n,\, \vert\,L\,\vert+1,\,\phi_a \rho^2\,\right)\,,
\end{equation}
$$
    \int\limits_0^\infty \rho\, d\rho\, \left( u_{n L} u_{m L} \right) = \delta_{nm}.
$$
The maximal eigenvalue $\tau$ of the above problem determines the
shift in the critical temperature of the FFLO transition. Within
the first-order perturbation theory in $v_0$ one can  get the
following expression for the temperature shift $\tau_{n L}$  vs
the dimensionless rotation frequency $\phi_a$:
\begin{equation} \label{eq:15g}
   \tau_{n L} = \tau_{n L}^{(0)} - v_{n n}^L \ .
\end{equation}
Thus, perturbation theory provides us a simple estimate for the
FFLO transition temperature:
$$
\tau = \underset{L\ge 0}{\rm max}
\left[(4\phi_a-v_0/\phi_a)(2L+1)+v_0L/\phi_a-4\phi_a^2(2L+1)^2
\right]$$

In Fig.~5  we show the results of the numerical calculation of the
dependencies $\tau(\phi_a)$ for different trapping frequencies.
These phase diagrams appear to be in good qualitative agreement
with the above estimate for not too small $\phi_a$ values. For
rather large trapping frequencies one can clearly observe the
regime of the rotation induced superfluid transition.

%%%%%%%%%%%%%%%%%%%%%%%%%%%%%%%%%%%%%%%%%%%%%%%%%%%%%%%%%%%%%%%%%%%%

\begin{figure}[t]
\includegraphics[width=0.55\textwidth]{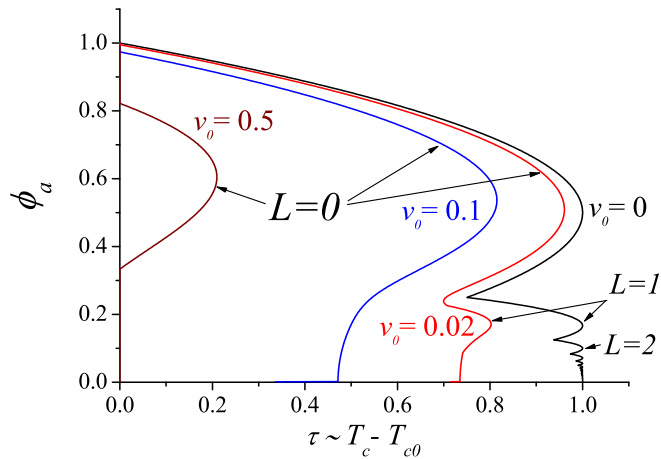}
\caption{(Color online) Typical phase diagrams in the plane
$(\tau,\phi_a)$ of a rotating Fermi condensate for different
trapping frequencies ($v_0 = 0;\, 0.02;\, 0.1;\, 0.5$). The arrows
point to the segments of the $\tau(\phi_a)$ curves corresponding
to different values of  vorticity $L=0, 1, 2$.}
\end{figure}
%
%%%%%%%%%%%%%%%%%%%%%%%%%%%%%%%%%%%%%%%%%%%%%%%%%%%%%%%%%%%%%%%%%%%%

\section{Conclusions}
To sum up, we have studied the effect of confinement of
superconducting and superfluid condensates on the phenomenon of
FFLO instability. We have found the following hallmarks of the
FFLO phenomenon in a restricted geometry: (i) both the finite
system size and parabolic trapping potential are responsible for
suppression of the quantum oscillations of the superfluid critical
temperature; (ii) the spatial oscillations of the superfluid order
parameter in the FFLO regime are suppressed by the increase in the
trapping frequency; (iii) change in the Zeeman interaction energy
in the mesoscopic superconducting system can induce phase
transitions between different inhomogeneous FFLO states; (iv)
switching between the vortex states in confined geometry can be
accompanied by giant jumps in vorticities; (v) rotation induced
superfluid transition in a Fermi gas cloud for rather large
trapping frequency; (vi) superconducting transition induced by the
perpendicular magnetic field component in a mesoscopic
superconducting disc; (vii) unusual scaling in the dependence of
the field of the vortex entry vs system size in the vicinity of
FFLO instability. We believe that  these theoretical predictions
can be used for experimental identification of the FFLO phases in
both mesoscopic superconductors and superfluid Fermi gases. Note
in conclusion that the physics of the vorticity switching in the
systems studied in this paper is in some respects similar to the
switching phenomena in multiply connected hybrid
superconductor/ferromagnet structures where the imbalance in spin
populations is induced by the ferromagnet exchange field
\cite{Samokhvalov-FS}.

\acknowledgments
 This work was supported, in part, by the Russian Foundation for
Basic Research,  Russian Agency of Education under the Federal
Program "Scientific and educational personnel of innovative Russia
in 2009-2013",   International Exchange Program of Universite
Bordeaux I, by the French ANR program NT09-612693 "SINUS", by the
"Dynasty" Foundation, and by the program of LEA Physique Theorique
et Matiere Condensee.

\end{document}